\documentclass[envcountsame,runningheads]{llncs}

\usepackage{amsmath}
\usepackage{amssymb}
\usepackage{latexsym} 
\usepackage[all]{xypic}
\usepackage{comment}
\usepackage{ifthen}
\usepackage{hyperref}

\def\finalversion{}
\ifthenelse{\isundefined{\finalversion}}{%
   \excludecomment{finalversion}%
   \includecomment{draftversion}%
}{
  \includecomment{finalversion}%
  \excludecomment{draftversion}%
}

\ifthenelse{\isundefined{\longversion}}{%
  \excludecomment{longversion}%
  \includecomment{shortversion}%
}{
  \excludecomment{shortversion}%
  \includecomment{longversion}
}

\excludecomment{trash}

\input macros.tex

\title{A Nice Labelling for Tree-Like \\ Event Structures of Degree $3$}
\author{Luigi Santocanale}%
\institute{%
  Laboratoire d'Informatique Fondamentale de Marseille,
  Universit\'e de Provence, 
  \\39 rue F. Joliot Curie, 13453 Marseille Cedex 13, France
  \\ \email{luigi.santocanale@lif.univ-mrs.fr}
}

\titlerunning{Nice Labelling for Tree-Like Event Structures}

\begin{document}

\maketitle

\begin{abstract}
  We address the problem of finding nice labellings for event
  structures of degree $3$. We develop a minimum theory by which we
  prove that the labelling number of an event structure of degree $3$
  is bounded by a linear function of the height.  The main theorem we
  present in this paper states that event structures of degree $3$
  whose causality order is a tree have a nice labelling with $3$
  colors.  Finally, we exemplify how to use this theorem to construct
  upper bounds for the labelling number of other event structures of
  degree $3$.
\end{abstract}


\section{Introduction}

\shtbs{Introduce event structures and why they are important to
  concurrency}

Event structures, introduced in \cite{eventstructures1}, are nowadays
a widely recognized model of true concurrent computation and have
found many uses since then. They are an intermediate abstract model
that make it possible to relate other more concrete models such as
Petri Nets or higher dimensional 
automata \cite{winskelnielsen}. They
provide formal semantics of process calculi \cite{winskelccs,varacca}.
More recently, logicians became interested in event structures with the
aim of constructing models of proof systems that are invariant under
the equalities induced by the cut elimination procedure
\cite{faggian,mellies}.

\shtbs{What is an event structure, the nice labelling problem}

Our interest for event structures stems from the fact that they
combine distinct approaches to the modeling of concurrent computation.
On one side, language theorists have developed the theory of partially
commutative monoids \cite{bookoftraces} as the basic language to
approach concurrency. On the other hand, the framework of domain
theory and, ultimately, order theoretic ideas have often been proposed
as the proper tools to handle concurrency, see for example
\cite{pratt}.  In this paper we pursue a combinatorial problem that
lies at the intersection of these two approaches.  It is the problem
of finding nice labellings for event structures of fixed degree. To
our knowledge, this problem has not been investigated any longer since
it was posed and partially solved in \cite{rozoy}.

\shtbs{The nice labelling problem} 

Let us recall that an event structure is made up of a set of local
events $E$ which is ordered by a causality relation $\leq$. Moreover,
a concurrency relation $\conc$, that may only relate causally
independent events, is given. A global state of the computation is
modeled as a clique of the concurrency relation. Global states may be
organized into a poset, the coherent domain of an event structure,
which represents all the concurrent non-deterministic executions of a
system.  Roughly speaking, the nice labelling problem consists in
representing the coherent domain of an event structure as a poset of
traces or, more precisely, pomsets.  That is, such a domain should be
reconstructed using the standard ingredients of trace theory: an
alphabet $\Sigma$, a local independence relation $I$, and a prefix
closed subset of the free monoid $L$, see \cite{arnold,thiagarajan}.
By the general theory relating traces to ordered sets, the problem
always has a solution $(\Sigma,I,L)$. We are asked to find a solution
with the cardinality of the alphabet $\Sigma$ minimal.  The problem is
actually equivalent to a graph coloring problem in that we can
associate to an event structure a graph, of which we are asked to
compute the chromatic number.
The degree of an event structure is the maximal number of upper covers
of some elements in the associated domain. Under the graph theoretic
translation of the problem, the degree coincides with the clique
number, and therefore it is a lower bound for the cardinality of a
solution.  A main contribution in \cite{rozoy} was to prove that event
structures of degree $2$ have a nice labelling with $2$ letters, i.e.
they have a solution $(\Sigma,I,L)$ with $\card(\Sigma) = 2$. On the
other hand, it was proved there that event structures of higher
degrees may require more letters than the degree.

The labelling problem may be thought to be a generalization of the
problem of covering a poset by disjoint chains.  Dilworth's Theorem
\cite{dilworth} states that the minimal cardinality of such a cover
equals the maximal cardinality of an antichain. This theorem and the
results of \cite{rozoy} constitute the few knowledge on the problem
presently available to us. For example, we cannot state that there is
some fixed $k > n$ for which every event structure of degree $n$ has a
nice labelling with at most $k$ letters.  In light of standard graph
theoretic results \cite{mycielski}, the above statement should not be
taken for granted.

\shtbs{Contribution here}

We present here our first results on the nice labelling problem for
event structures of degree $3$.  We develop a minimum theory that
shows that the graph of a degree $3$ event structure, when restricted
to an antichain, is almost acyclic and can be colored with $3$
letters.  This observation allows to construct an upper bound to the
labelling number of such event structure as a linear function of its
height.  We prove then our main theorem stating that event structures
of degree $3$, whose causality order is a tree, have a nice labelling
with $3$ letters.  Let us just say that such an event structure may
represent a concurrent system where some processes are only allowed to
fork or to take local nondeterministic choices.  Finally, we exemplify
how to use this theorem to construct upper bounds for the labelling
number of other event structures of degree $3$. In some simple cases,
we obtain constant upper bounds to the labelling number, i.e. upper
bounds that are functions of no parameter.

While these results do not answer the general problem, that of
computing the labelling number of degree $3$ event structures, we are
aware that graph coloring problems may be difficult to answer. Thus we
decided to present these results and share the knowledge so far
acquired and  also  to encourage other researchers to pursue
the problem.
%
%
Let us mention why we believe that this and other problems in the
combinatorics of concurrency deserve to be deeply investigated. The
theory of event structures is now being applied within verification.
A model checker, POEM, presently developed in Marseilles, makes
explicit use of trace theory and of the theory of partial orders to
represent the state space of a concurrent system \cite{poem}. The
combinatorics of posets is there exploited to achieve an the efficient
exploration of the global states of concurrent systems
\cite{partialorderreduction}.  Thus, having a solid theoretical
understanding of such combinatorics is, for us, a prerequisite and a
complement for designing efficient algorithms for these kind of tools.

\shtbs{Structure of the paper}

The paper is structured as follows. After recalling the order
theoretic concepts we shall use, we introduce event structures and the
nice labelling problem  in section \ref{sec:theproblem}.
In section \ref{sec:antichains} we develop the first properties of
event structures of degree $3$. As a result, we devise an upper bound
for the labelling number of such event structures as a linear function
of the height.
In section \ref{sec:thetheorem} we present our main result stating
that event structures whose underlying order is a tree may be
labeled with $3$ colors.
In section \ref{sec:w2} we develop a general approach to construct
upper bounds to the labelling number of event structures of degree
$3$.  Using this approach and the results of the previous section, we
compute a constant upper bound for a class of degree $3$ event
structures that have some simplifying properties and which are
consequently called simple.


\subsubsection*{Order Theoretic Preliminaries.}

We shall introduce event structures in the next section. For the
moment being let us anticipate that part of an event structure is a
set $E$ of events which is partially ordered by a causality relation
$\leq$. In this paper we shall heavily make use of order theoretic
concepts. We introduce them here together with the notation
that shall be used.  All these concepts will apply to the poset
$\langle E,\leq\rangle$ of an event structure.
 
A finite poset is a pair $\langle P,\leq \rangle$ where $P$ is a
finite set and $\leq$ is a reflexive, transitive and antisymmetric
relation on $P$.  A subset $X \subseteq P$ is a \emph{lower set} if $y
\leq x \in X$ implies $y \in X$. If $Y \subseteq P$, then we denote by
$\idealof{Y}$ the least lower set containing $Y$. The explicit formula
for $\idealof{Y}$ is
\begin{align*}
  \idealof{Y} & =\set{x \in P\,|\,\exists y \in Y \tst x \leq y}\,.
\end{align*}
Two elements $x,y \in P$ are \emph{comparable} if and only if either
$x \leq y$ or $y \leq x$. We write $x \comp y$ to mean that $x,y$ are
comparable.  A \emph{chain} is sequence $x_{0},\ldots ,x_{n}$ of
elements of $P$ such that $x_{0} < x_{1} < \ldots < x_{n}$. The
integer $n$ is the length of the chain.  
The \emph{height} of an
element $x \in P$, noted $\height(x)$, is the length of the longest
chain in $\ideal{x}$. The height of $P$ is $\max \set{\height(x)\,|\,
  x \in P}$.  
An \emph{antichain} is a subset $X \subseteq P$ such that $x \not\comp
y$ for each pair $x,y \in X$.  The \emph{width} of $\langle P, \leq
\rangle$, noted $\width(P,\leq)$, is the integer $\max
\set{\card(A)\,|\, A \text{ is an antichain}}$.
If the interval $\set{z \in P\,|\, x \leq z \leq y}$ is the two
elements set $\set{x,y}$, then we say that $x$ is a \emph{lower cover}
of $y$ or that $y$ is \emph{an upper cover} of $x$. We denote this
relation by $x \lcover y$. The Hasse diagram of $\langle P,\leq
\rangle$ is the directed graph $\langle P,\lcover \rangle$.  For $x
\in P$, the \emph{degree} of $x$, noted $\deg(x)$, is the number of
upper covers of $x$. That is, the degree of $x$ is the outdegree of
$x$ in the Hasse diagram. The degree of $\langle P, \leq \rangle$,
noted $\deg(P,\leq)$, is the integer $\max \set{\deg(x)\,|\,x \in P}$.
We shall denote by $\fat(x)$ the number of lower covers of $x$ (i.e.
the indegree of $x$ in the Hasse diagram).  The poset $\langle P,\leq
\rangle$ is \emph{graded} if $x \lcover y$ implies $\height(y) =
\height(y) + 1$.

\vfill

 
\section{Event Structures and the Nice Labelling Problem}
\label{sec:theproblem}

Event structures are a basic model of concurrency introduced in
\cite{eventstructures1}. The definition we present here is from
\cite{winskelnielsen}.
\begin{definition}
  An \emph{event structure} is a triple $\E = \langle
  E,\leq,\CNF\rangle$ such that
  \begin{itemize}
  \item $\langle E,\leq\rangle$ is a poset, such that for each $x \in
    E$ the lower set $\ideal{x}$ is finite,
  \item $\CNF$ is a collection of subsets of $E$ such that:
    \begin{enumerate}
    \item $\set{x} \in \CNF$ for each $x \in E$,
    \item $X \subseteq Y \in \CNF$ implies $X \in \CNF$,
    \item $X \in
      \CNF$ implies $\idealof{X} \in \CNF$.
    \end{enumerate}
  \end{itemize}
\end{definition}
The order $\leq$ of an event structure $\E$ is known as the
\emph{causality} relation between events. The collection $\CNF$ is
known as the set of configurations of $\E$. A configuration $X \in
\CNF$ of causally unrelated events -- that is, an antichain w.r.t.
$\leq$ -- is a sort of snapshot of the global state of some
distributed computation. A snapshot $X$ may be transformed into a
description of the computation that takes into account its history.
This is done by adding to $X$ the events that causally have determined
events in $X$. That is, the history aware description is the lower set
$\idealof{X}$ generated by $X$.

Two elements $x,y \in E$ are said to be \emph{concurrent} if $x
\uncomp y$ and there exists $X \in \CNF$ such that $x,y \in X$.  Two
concurrent elements will be thereby noted by $x \conc y$.  It is
useful to introduce a weakened version of the concurrency relation
where we allow elements to be comparable: $x \wconc y$ if and only if
$x \conc y$ or $x \comp y$. Equivalently, $x \wconc y$ if and only if
there exists $X \in \CNF$ such that $x,y \in X$.
In many concrete models the set of configurations is completely
determined by the concurrency relation.
\begin{definition}
  An event structure $\E$ is \emph{coherent} if $\CNF$ is the set
  of cliques of the weak concurrency relation: $X \in \CNF$ if
  and only if $x \wconc y$ for every pair of elements $x,y \in X$.
\end{definition}
Coherent event structures are also known as event structures with
binary conflict. To understand the naming let us explicitely introduce
the conflict relation and two other derived relations:
\begin{itemize}
\item \emph{Conflict}: $x \confl y$ if and only if $x \ncomp y$ and $x
  \nconc y$.
\item \emph{Minimal conflict}: $x \mconfl y$ if and only (i) $x \confl
  y$, (ii) $x' < x$ implies $x' \wconc y$, and (iii) $y' < y$ implies
  $x \wconc y'$.
\item \emph{Orthogonality}: $x \orth y$ if and only if $x \mconfl y$
  or $x \conc y$.
\end{itemize}
A coherent event structure is completely described by a triple
$\langle E,\leq,\conc\rangle$ where the latter is a symmetric relation
subject to the following condition: $x \conc y$ and $z \leq x$ implies
$z \conc y$ or $z \leq y$.  Similarly, a coherent event structure is
completely determined by the order and the conflict relation.  In this
paper we shall deal with coherent event structures only and, from now
on, event structure will be a synonym for coherent event structure.

\breath

We shall focus mainly on the orthogonality relation. Let us observe
that two orthogonal elements are called independent in \cite{rozoy}.
We prefer however not to use this naming: we shall frequently make use
of standard graph theoretic language and argue about cliques, not on
their dual, independent sets.  The orthogonality relation clearly is
symmetric and moreover it inherits from the concurrency relation the
following property: if $x \orth y$ and $z \leq x$, then $z \orth y$ or
$z \leq y$.
\begin{definition}
  A \emph{nice labelling} of an event structure $\E$ is a pair
  $(\lambda,\Sigma)$, where $\Sigma$ is a finite alphabet and $\lambda
  : E \rTo \Sigma$ is such that $\lambda(x) \neq \lambda(y)$ whenever
  $x \orth y$.
\end{definition}
That is, if we let $\G(\E)$ -- the graph of $\E$ -- be the pair
$\langle E, \orth \rangle$, a labelling of $\E$ is a coloring of the
graph $\G(\E)$.  For a graph $G$, let $\chr(G)$ denote its chromatic
number. Let us say then that the \emph{labelling number} of $\E$ is
$\chr(\G(\E))$.  The \emph{nice labelling problem} for a class
$\mathcal{K}$ of event structures amounts to computing the number
\begin{align*}
  \chr(\mathcal{K}) & = \max \set{\chr(\G(\E))\,|\,\E \in
    \mathcal{K}}\,.
\end{align*}
To understand
the origins of this problem, let us recall the definition of the
domain of an event structure.
\begin{definition}
  The domain $\D(\E)$ of an event structure $\E = \langle E,\leq,\CNF
  \rangle$ is the collection of lower sets in $\CNF$, ordered by
  subset inclusion.
\end{definition}
Following a standard axiomatization in theoretical computer science
\cite{winskelnielsen} $\D(\E)$ is a stable domain which is
coherent if $\E$ is coherent.  Stable means that $\D(\E)$ is
essentially a distributive lattice. As a matter of fact, if $\E$ is
finite, then a possible alternative axiomatization of the poset
$\D(\E)$ is as follows.  It is easily seen that the collection
$\mathcal{D}(\E)$ is closed under binary intersections, hence it is a
finite meet semilattice without a top element, a chopped lattice in
the sense of \cite[Chapter 4]{gratzer}.  Also the chopped lattice is
distributive, meaning that whenever $X,Y,Z \in \D(\E)$ and $X \cup Y
\in \D(\E)$, then $Z \cap (X \cup Y) = (Z \cap X) \cup (Z \cap Y)$. It
can be shown that every distributive chopped lattice is isomorphic to
the domain of a finite -- not necessarily coherent -- event structure.
\begin{longversion}
  \begin{proposition}
    Every distributive chopped lattice is isomorphic to the stable
    domain of an event structure.
  \end{proposition}
  \begin{proof}
    First of all we remark that it makes sense requiring that a
    chopped lattice is distributive. As a matter of fact, in a chopped
    lattice the join of $x,y$ is uniquely defined as soon as the set
    of upped bounds of $x,y$ is not empty. Since $(z \land x), (z
    \land y)\leq z \land (x \vee y)$, existence of $x \vee y$ implies
    the existence of $(z \land x)\vee (z \land y)$. Hence the
    distributivity conditions, stating that $x \vee y$ exists implies
    $(z \land x)\vee (z \land y)= z \land (x \vee y)$ makes sense for
    any chopped lattice.

    If $L$ is a distributive chopped lattice, say that $x \in L$ is
    prime if it has a unique lower cover. It is standard to argue that
    $x \leq z \vee y$ implies $x \leq z$ or $x \leq y$. Denote by
    $J(L)$ the primes of $L$. For $X \subseteq J(L)$, let us say that
    $X \in \CNF$ if $\bigvee X$ exists, and let $\E = \langle
    J(L),\leq,\CNF$.  It is a standard exercice to prove that $\D(\E)$
    is order isomorphic to $L$.  \qed
  \end{proof}
\end{longversion}

\begin{lemma}
  \label{lemma:degree}
  A set $\set{x_{1},\ldots ,x_{n}}$ is a clique in the graph $\G(\E)$
  iff there exists $I \in \D(\E)$ such that $I \cup \set{x_{i}} $, $i
  = 1,\ldots ,n$, are distinct upper covers of $I$ in the domain
  $\D(\E)$.
\end{lemma}
\begin{longversion}
  \begin{proof}
    Suppose that $\set{x_{i}} \cup I$ and $\set{x_{j}} \cup I$ are
    distinct upper covers of some $I$ in $\mathcal{D}(\mathcal{E})$.
    Then $\{x_{i},x_{j}\}$ is an antichain since $x_{i} \leq x_{j}$
    implies that $\set{x_{i}} \cup I \leq \set{x_{j}} \cup I$.  If $x'
    < x_{i}$ then $I \subseteq \set{x'} \cup I \subseteq \set{x_{j}}
    \cup I$ hence $I = \set{x'} \cup I$ and $x' \in I \subseteq
    \set{x_{j}} \cup I$. Since $\set{x_{j}} \cup I$ is a
    $\concurrent$-clique, then $x' \concurrent x_{j}$.  Similarly $y'
    < x_{j}$ implies $x_{i} \concurrent y'$. Thus, $\set{x_{1},\ldots
      ,x_{n}}$ is a clique.

    Conversely, let us suppose that $x_{i} \orthogonal x_{j}$ whenever
    $i \neq j$. Observe that $x_{i} \orthogonal x_{j}$ implies $x'
    \concurrent y'$ for $x' < x$ and $y'< y$.  Thus the ideal $I =
    \bigcup_{i=1}^{n}\set{x'\,|\,x' < x_{i}}$ belongs to
    $CL(\mathcal{E})$ and $\set{x_{i}} \cup I$ belongs to
    $CL(\mathcal{E})$ for $i =1,\ldots ,n$. Since $\set{x_{i},x_{j}}$
    is an antichain, then $\set{x_{i}} \cup I$ and $\set{x_{j}} \cup
    I$ are distinct upper covers of $I$.
  \end{proof}
\end{longversion}
The Lemma shows that a nice labelling $\lambda : E \rTo \Sigma$ allows
to label the edges of the Hasse diagram of $\D(\E)$ so that: (i)
outgoing edges from the same source vertex have distinct labels, (ii)
perspective edges -- i.e. edges $I_{0} \lcover I_{1}$ and $J_{0}
\lcover J_{1}$ such that $I_{0} = I_{1} \cap J_{0}$ and $J_{1} = I_{1}
\cup J_{0}$ -- have the same label. These conditions are necessary and
sufficient to show that $\D(\E)$ is order isomorphic to a consistent
(but not complete) set of $P$-traces (or pomsets) on the alphabet
$\Sigma$ in the sense of \cite{arnold}.

The \emph{degree} of an event structure $\E$ is the degree of the
domain $\D(\E)$, that is, the maximum number of upper covers of a
lower set $I$ within the poset $\D(\E)$. Lemma \ref{lemma:degree}
shows that the degree of $\E$ is equal to the size of a maximal clique
in $\G(\E)$, i.e. to the clique number of $\G(\E)$. Henceforth, the
degree of $\E$ is a lower bound to $\chr(\G(\E))$.  The following
Theorems state the few results on the nice labelling problem that are
available in the literature.
\begin{theorem}[see \cite{dilworth}]
  Let $\mathcal{NC}_{n}$ be the class of event structures of degree at
  most $n$ with empty conflict relation. Then $\chr(\mathcal{NC}_{n})
  = n$.
\end{theorem}

\begin{theorem}[see \cite{rozoy}]
  Let $\mathcal{K}_{n}$ be the class of event structures of degree at
  most $n$. Then $\chr(\mathcal{K}_{n}) = n$ if $n \leq 2$ and
  $\chr(\mathcal{K}_{n}) \geq n + 1$ otherwise.
\end{theorem}
The last theorem has been our starting point for investigating the
nice labelling problem for event structures of degree $3$.

\begin{longversion}
  \marginpar{Is this correct ???}  The reader may have noticed the
  similarity between the nice labelling problem for the class
  $\mathcal{K}_{n}$ of event structure of fixed degree, and the
  problem of finding an upper bound to the chromatic number of graphs
  of fixed clique number. Since the latter has a negative solution, it
  might be conjectured that also the nice labelling problem for the
  classes $\mathcal{K}_{n}$ might have a negative solution. The graphs
  of the form $\G(\E)$ are subject to many constraints, for example
  minimal elements of $\E$ form a clique in $\G(\E)$. It is possible
  to show that many well known families of graphs do not occur as
  subgraphs of some $\G(\E)$ with $\E \in \mathcal{K}_{n}$. The
  following Proposition illustrates this point.

\begin{proposition}
  Let $\mathcal{M}$ be the Mycielski transform of a graph
  \cite{mycielski} and let $K_{3}$ the total graph on three elements.

  The Mycielski graphs $M^{n}(K_{3})$, $n \geq 2$, do not occur as a
  subgraphs of some graph of the form $\mathcal{G}(\mathcal{E})$ such
  that $\cl(\mathcal{G}(\mathcal{E})) = 3$.
\end{proposition}
\end{longversion}

\begin{longversion}
  \subsubsection{Characterization of data of the form $\langle
    P,\leq,\orthogonal\rangle$.}

  The following observations are trivial but worth to remark.

  Let us consider a triple $\langle P,\leq,E \rangle$, where $\langle
  P,E\rangle$ is an undirected graph and $\langle P,\leq\rangle$ is a
  partially ordered set. Let us say that this triple is \emph{order
    consistent} if $x E y$ implies that (i) $x,y$ are uncomparable and
  (ii) $x ' < x$ implies $x' E y$ or $x' \leq y$.  If the triple is
  order consistent, then we say that a pair $(x,y)$ is \emph{left
    induced} if either $x \leq y$, or $x'Ey$ for some $x' > x$; it is
  \emph{right induced} if either $x \geq y$, or $xEy'$ for some $y' >
  y$. A pair $(x,y)$ is \emph{$E$-induced} iff it is left or right
  induced.  We define $E^{\circ}$ to be the set of $E$-induced
  pairs.\footnote{%
    Let $Compl = \leq \cup \geq$, then $Comp \subseteq E^{\circ}
    \subseteq E \cup Comp$.  } We say that an order consistent $E$ is
  \emph{closed} if whenever $x$ and $y$ are such that $x'E^{\circ} y$
  for every $x' <x$ and $x E^{\circ} y'$ for every $y' < y$, then $x E
  y$.

\begin{proposition}
  The triple $\langle P,\leq,\orthogonal\rangle$ is closed order
  consistent.  Conversely, if a triple $\langle P,\leq,E \rangle$ is
  closed order consistent then $E$ is of the form $\orthogonal$ for
  the concurrecny relation $E^{\circ}$; the latter is the least
  concurrency relation giving rise to $E$.
\end{proposition}
\begin{proof}
  It is straightforward to observe that $\orthogonal$ is closed and
  consistent, since if $x \orthogonal^{\circ} y$, then $x \concurrent
  y$.

  Let us consider $E$ which is closed consistent. Let us define
  $\concurrent$ as $E^{\circ}$, then closedness of $E$ means that $x
  \orthogonal y$ implies $x E y$. Conversely, let us suppose that $xE
  y$. Then $x,y$ is an antichain and if $x' < y$, then $x' E^{\circ}
  y$, and similarly for $y$: thus $x \orthogonal y$.

  Finally, if $\concurrent$ gives rise to $E$, and $x E^{\circ} y$,
  then if $x$ and $y$ are comparable, then $x \concurrent y$,
  otherwise $x' E y$ for some $x' > x$ (or the symmetric case) then $x
  \concurrent yE$, showing that $E \subseteq \concurrent$.
  
\end{proof}

\begin{proposition}
  Consider a closed order consistent $\langle
  P,\leq,\orthogonal\rangle$ triple and let $p \in P$. Let $P_{p} =
  \set{p'\,|\,pEp'}$. Then $\langle
  P_{p},\leq_{|P_{p}},E_{|P_{p}}\rangle$ id closed order consistent.
\end{proposition}
\begin{proof}
  Clearly, the triple is order consistent. Thus suppose that for each
  $x, y \in P_{p}$, and suppose that $x' \in P_{p}$ and $x' < x$
  implies $x' E^{\circ} y$ and the similar relation holds for $y$.

  Let $x' < x$ with $x' \not\in P_{p}$. Since $E$ is closed, $x'< p$
  and thus $x'E^{\circ} y$.

  Therefore $x,y$ is a critical pair, and therefore $x E y$ since $E$
  is closed.  
\end{proof}
\end{longversion}


\section{Cycles and Antichains}
\label{sec:antichains}

From now own, in this and the following sections, $\E = \langle E,
\leq, \CNF\rangle$ will denote a coherent event structure of degree at
most $3$.  We begin our investigation of the nice labelling problem by
studying the restriction to an antichain of the graph $\G(\E)$.
The main tool we shall use is the following Lemma. It is a
straightforward generalization of \cite[Lemma 2.2]{rozoy} to degree
$3$. In \cite{getco06} we proposed generalizations of this Lemma to
higher degrees, pointing out their strong geometrical flavor.
\begin{lemma}
  \label{lemma:avoided}
  Let $\set{x_{0},x_{1},x_{2}}, \set{x_{1},x_{2},x_{3}}$ be two size
  $3$ cliques in the graph $\G(\E)$ sharing the same face
  $\set{x_{1},x_{2}}$. Then $x_{0},x_{3}$ are comparable.
\end{lemma}
\begin{proof}
  Let us suppose that $x_{0},x_{3}$ are not comparable. It is not
  possible that $x_{0} \orth x_{3}$, since then we have a size $4$
  clique in the graph $\G(\E)$. Thus $x_{0} \confl x_{3}$ and we can
  find  $x'_{0}
  \leq x_{0}$ and $x'_{3} \leq x_{3}$ such that $x'_{0} \mconfl
  x'_{3}$.  We claim that $\set{x'_{0},x_{1},x_{2},x'_{3}}$ is a size
  $4$ clique in $\G(\E)$, thus reaching a contradiction.

  If $x'_{0} \north x_{1}$, then $x'_{0} \leq x_{1}$, but this,
  together with $x_{1} \orth x_{3}$, contradicts $x'_{0} \mconfl
  x'_{3}$. Similalry, $x'_{0} \orth x_{2}$, $x'_{3} \orth x_{1}$,
  $x'_{3}\orth x_{2}$.
  \qed
\end{proof}

We are going to improve on the previous Lemma. To this goal, let us
say that a sequence $x_{0}x_{1}\ldots x_{n-1}x_{n}$ is a
\emph{straight cycle} if $x_{n} = x_{0}$, $x_{i} \orth x_{i +1}$ for
$i = 0,\ldots ,n-1$, $x_{i} \not\comp x_{j}$ whenever $i, j \in
\set{0,\ldots ,n-1}$ and $i \neq j$. As usual, the integer $n$ is the
length of the cycle.  Observe that a straight cycle is simple, i.e., a
part from the endpoints of the cycle, it does not visit twice the same
vertex . The height of a straight cycle $C = x_{0}x_{1}\ldots x_{n}$
is the integer
\begin{align*}
  \heightp(C) = \sum_{i = 0,\ldots ,n-1} \height(x_{i}) + 1\,. 
\end{align*}
The definition of $\heightp$ implies that if $C'$ is a subcycle of $C$ induced
by a chord, then $\heightp(C') < \heightp(C)$.

\begin{proposition}
  The graph $\G(\E)$ does not contain a straight cycle of length
  strictly greater than $3$.
\end{proposition}
\begin{proof}
  Let $\CYf$ be the collection of straight cycles in $\G(\E)$ whose
  length is at least $4$. We shall show that if $C \in \CYf$, then
  there exists $C' \in \CYf$ such that $\heightp(C') < \heightp(C)$.

  Let $C$ be the straight cycle $x_{0}\orth x_{1}\orth x_{2}\ldots
  x_{n-1} \orth x_{n} = x_{0}$ where $n \geq 4$. Let us suppose that
  this cycle has a chord. It follows, by Lemma \ref{lemma:avoided},
  that $n > 4$. Hence the chord divides the cycle
  into two straight cycles, one of which has still length at least $4$
  and whose height is less than the height of $C$, since it
  contains a smaller number of vertices.

  Otherwise $C$ has no chord and $x_{0}\north x_{2}$. This means that
  either there exists $x'_{0} < x_{0}$ such that $x'_{0}\nconc x_{0}$,
  or there exists $x'_{2} < x_{2}$ such that $x_{0}\nconc x'_{2}$.  By
  symmetry, we can assume the first case holds. As in the proof of
  Lemma \ref{lemma:avoided} $\set{x'_{0},x_{1},x_{2},x_{3}}$ form an
  antichain, and $x'_{0}x_{1}x_{2}x_{3}$ is a path.  Let $C'$ be the
  set $\set{x'_{0}x_{1},\ldots x_{n-1}x_{0}'}$. If $C'$ is an
  antichain, then $C'$ is a straight cycle such that $\heightp(C') <
  \heightp(C)$.  Otherwise the set $\set{j\in \set{4,\ldots
      ,n-1}\,|\,x_{j} \geq x'_{0}}$ is not empty.  Let $i$ be the
  minimum in this set, and observe that $x_{i-1}\orth x_{i}$ and
  $x'_{0} \leq x_{i}$ but $x'_{0} \not\leq x_{i-1}$ implies
  $x_{i-1}\orth x'_{0}$.  Thus $\tilde{C} =
  x'_{0}x_{1}x_{2}x_{3}\ldots x_{i-1} x'_{0}$ is a straight cycle of
  lenght at least $4$ such that $\heightp(\tilde{C}) < \heightp(C)$.
  \qed
\end{proof}

\begin{corollary}
  \label{cor:antichain}
  Any subgraph of $\G(\E)$ induced by an antichain can be colored with
  $3$ colors.
\end{corollary}
\begin{proof}
  Since the only cycles have length at most $3$, such an induced graph
  is chordal and its clique number is $3$. It is well known that the
  chromatic number of chordal graphs equals their clique number
  \cite{fulkerson}.  \qed
\end{proof}

In the rest of this section we exploit the previous observations to
construct upper bounds for the labelling number of $\E$. We remark that
these upper bounds might appear either too abstract, or too trivial.
On the other hand, we believe that they well illustrate the kind of
problems that arise when trying to build complex event structures that
might have labelling number greater than $4$.

A \emph{stratifying function} for $\E$ is a function $h : E \rTo \N$
such that, for each $n \geq 0$, the set $\set{x \in E\,|\,h(x) = n}$
is an antichain. The height function is a stratifying function. Also
$\varsigma(x) = \card{\set{y \in E\,|\,y < x}}$
is a stratifying function.  With respect to a stratifying function $h$
the $h$-skewness of $\E$ is defined by
\begin{align*}
  \skewn[h](\E) & = \max\set{|h(x) - h(y)|\,|\, x\orth y}\,.
  \intertext{More generally, the skewness of $\E$ is defined by}
  \skewn(\E) & = \min\set{%
    \skewn[h](\E)\,|\, h \text{ is a stratifying function }}\,.
\end{align*}

\begin{proposition}
  \label{prop:skew}
  If $\skewn(\E) < n$ then $\chr(\G(\E)) \leq 3n$.
\end{proposition}
\begin{proof}
  Let $h$ be a stratifying function such that $|h(x) - h(y)| < n$
  whenever $x \orth y$.  For each $k \geq 0$, let $\lambda_{k} :
  \set{x\in E \,|\,h(x) = k} \rTo \set{a,b,c}$ be a coloring of the
  graph induced by $\set{x\in E \,|\,h(x) = k}$. Define $\lambda :
  E \rTo \set{a,b,c}\times \set{0,\ldots ,n -1}$ as follows:
  \begin{align*}
    \lambda(x) & = (\lambda_{h(x)}(x), h(x)\mymod n)\,.
  \end{align*}
  Let us suppose that $x \orth y$ and $h(x) \geq h(y)$, so that $0
  \leq h(x) - h(y) < n$. If $h(x) = h(y)$, then by construction
  $\lambda_{h(x)}(x) = \lambda_{h(y)}(x)\neq \lambda_{h(y)}(y)$.
  Otherwise $h(x) > h(y)$ and $0 \leq h(x) - h(y) < n$ implies $h(x)
  \mymod n \neq h(y) \mymod n$. In both cases we obtain $\lambda(x)
  \neq \lambda(y)$.  \qed
\end{proof}

An immediate consequence of Proposition \ref{prop:skew} is the
following upper bound for the labelling number of $\E$:
\begin{align*}
   \gamma(\G(\E)) & \leq 3(\height(\E) + 1)\,.
\end{align*}
To appreciate the upper bound, consider that another approximation to
the labelling number of $\E$ is provided by Dilworth's Theorem
\cite{dilworth}, stating that $\gamma(\G(\E)) \leq \width(\E)$.
To compare the two bounds, consider that there exist event structures
of degree $3$ whose width is an exponential function of the height.


\section{An Optimal Nice Labelling for Trees and Forests}
\label{sec:thetheorem}

We prove in this section the main contribution of this paper. Assuming
that $\langle E,\leq\rangle$ is a tree or a forest, then we define a
labelling with $3$ colors, and prove it is a nice labelling.  Since
clearly we can construct a tree which needs at least three colors,
such a labelling is optimal. Before defining the labelling, we shall
develop a small amount of observations.

\begin{definition}
  We say that two distinct events are \emph{twins} if they have the
  same set of lower covers.
\end{definition}
Clearly if $x,y$ are twins, then $z < x$ if and only if $z < y$.  More
importantly, if $x,y$ are twins, then the relation $x \orth y$ holds.
As a matter of fact, if $x' < x$ then $x' < y$, hence $x' \wconc y$.
Similarly, if $y' < y$ then $y' \wconc x$. It follows that a set of
events having the same lower covers form a clique in $\G(\E)$, hence
it has at most the degree of an event structure, $3$ in the present
case. To introduce the next Lemmas, if $x \in Y \subseteq E$, define
\begin{align*}
  O^{Y}_{x} & = \set{z\,|\,z \orth x \tand y \not\leq z, \text{
      forall } y \in Y }\,.
\end{align*}
If $Y = \set{x,y}$, then we shall abuse of notation and write
$O^{x,y}_{x}$, $O^{y,x}_{x}$ as synonyms of $O^{Y}_{x}$. Thus $z \in
O^{x,y}_{x}$ if and only if $z \orth x$ and $y \not\leq z$.
\begin{lemma}
  \label{lemma:twinsthree}
  If $x,y,z$ are parwise distinct twins, then $O^{\{x,y,z\}}_{x} =
  \emptyset$.
\end{lemma}
\begin{proof}
  Let us suppose that $w \in O^{\{x,y,z\}}_{x}$. If $w \orth y$, then
  $w \comp z$ by Lemma \ref{lemma:avoided}. Since $z \not\leq w$, then
  $w < z$.  However this implies $w < x$, contradicting $w \orth x$.
  Hence $w \north y$ and we can find $w' \leq w$, $y' \leq y$ such
  that $w' \mconfl y'$. It cannot be the case that $y' < y$, otherwise
  $y' < x$ and the pair $(w',y')$, properly covered by the pair
  $(w,x)$, cannot be a minimal conflict. Thus $w' < w$, and $y'$
  equals to $y$.  We claim that $w' \in O^{\{x,y,z\}}$. As a matter of
  fact, $w'$ cannot be above any of the elements in $\set{x,y,z}$,
  otherwise $w$ would have the same property. From $w \orth x$ and $w'
  < w$, we deduce that $w' \orth x$ or $w' \leq x$. If the latter,
  then $w' < x$, so that $w' < y$, contradicting $w' \mconfl y$.
  Therefore $w' \orth x$ and $\set{w',x,y},\set{x,y,z}$ are two
  $3$-cliques sharing the same face $\set{x,y}$. As before, $w' \comp
  z$, leading to a contradiction. \qed
\end{proof}
\begin{lemma}
  \label{lemma:twins}
  If $x,y$ are twins, then $O^{x,y}_{x},O^{x,y}_{y}$ are comparable
  w.r.t. set inclusion and $O^{x,y}_{x} \cap O^{x,y}_{y}$ is a linear
  order.
\end{lemma}
\begin{proof}
  We observe first that if $z \in O^{x,y}_{x}$ and $w \in O^{x,y}_{y}$
  then $z \comp w$. As a consequence $O^{x,y}_{x} \cap O^{x,y}_{y}$ is
  linearly ordered.
  
  Let us suppose that there exists $z \in O^{x,y}_{x}$ and $w \in
  O^{x,y}_{y}$ such that $z \not\comp w$. Observe then that
  $\set{z,x,y,w}$ is an antichain: $y \not\leq z$, and $z < y$ implies
  $z < x$, which is not the case due to $z \orth x$. Thus $z \not\comp
  y$ and similarly $w \not\comp x$.

  Since there cannot be a length $4$ straight cycle, we deduce $z
  \north w$.  Let $z' \leq z$ and $w' \leq w$ be such that $z' \mconfl
  w'$.  We claim first that $z' \orth x$. Otherwise, $z' \leq x$ and
  $z' < x$, since $z' = x$ implies $x \leq z$. The relation $z' < x$
  in turn implies $z' < y$, which contradicts $z' \mconfl w'$.  Also
  it cannot be the case that $y \leq z'$, since otherwise $y \leq z$.
  Thus, we have argued that $z' \in O^{x,y}_{x}$. Similarly $w' \in
  O^{x,y}_{y}$. As before $\set{z',x,y,w'}$ is an antichain, hence
  $z',x,y,w'$ also form a length $4$ straight cycle, a
  contradiction.

  Observe now that $w \leq z \in O^{x,y}_{x}$ and $w \not\leq x$
  implies $w \in O^{x,y}_{x}$. From $w \leq z \orth x$ deduce $w \orth
  x$ or $w \leq x$.  Since $w \not\leq x$, then $w \orth x$. Also, if
  $y \leq w$ then $y \leq z$, which is not the case.

  Let $z \in O^{x,y}_{x} \setminus O^{x,y}_{y}$, pick any $w \in
  O^{x,y}_{y}$ and recall that $z,w$ are comparable.  We cannot have
  $z \leq w$ since $z \not\leq y$ implies then $z \in O^{x,y}_{y}$.
  Hence $w < z \in O^{x,y}_{x}$ and $w \not\leq x$ imply $w \in
  O^{x,y}_{x}$ by the previous observation.\qed
\end{proof}

The following Lemma will prove to be the key observation in defining
later a nice labelling.
\begin{lemma}
  \label{lemma:Oinclusion}
  Let $(x,y)$ $(z,w)$ be two pairs of pairwise distinct twins such
  that $z \in O_{x}^{x,y} \cap O_{y}^{x,y}$ and $w \not\leq x$. Then
  $O_{z}^{w,z} \supset O_{w}^{w,z}$.
\end{lemma}
\begin{proof}
  If $O_{z}^{w,z} \not\supset O_{w}^{w,z}$, then $O_{z}^{w,z}
  \subseteq O_{w}^{w,z}$ by Lemma \ref{lemma:twins}.
  Since $w \not\leq x$ and $w \neq y$, then $w \not\leq y$. We have
  shown that $x,y \in O_{z}^{w,z}$, hence $x,y \in O_{w}^{w,z}$.  It
  follows that $\set{x,y,z,w}$ is a size $4$ clique, a
  contradiction.\qed
\end{proof}


\breath

We come now to discuss some subsets of $E$ for which we shall prove
that there exists a nice labelling with $3$ letters. The intuitive
reason for that is the presence of many twins.
\begin{definition}
  A subset $T \subseteq E$ is a \emph{tree} if and only if
  \begin{itemize}
  \item each $x \in T$ has exactly one lower cover $\pi(x) \in E$,
  \item $T$ is convex: $x,z \in T$ and $x < y < z$ implies $y \in T$,
  \item if $x,y$ are minimal in $T$, then $\pi(x) = \pi(y)$.
  \end{itemize}
\end{definition}
If $T$ is a tree and $x \in T$, the height of $x$ in $T$, noted
$\height[T](x)$, is the cardinality of the set $\set{y \in T \,|\, y <
  x}$.  A linear ordering $\lord$ on $T$ is said to be compatible with
the height if it satisfies
\begin{align}
  \label{cond:height}
  \tag{HEIGHT}
  \height[T](x) < \height[T](y) & \timplies x \lord y\,.
\end{align}
It is not difficult to see that such a linear ordering always exists.
With respect to such linear ordering, define
\begin{align*}
  \O(x) & = \set{y \in T \,|\, y \orth x \tand y \lord x }\,,
  \;\;\;\;x \in T.
\end{align*} 
We shall represent $\O(x)$ as the disjoint union of $\C(x)$ and
$\L(x)$ where
\begin{align*}
  \C(x) & = \set{y \in \O(x)\,| \,z \lcover x \timplies z \leq y}\,,
  &
  \L(x) & = \O(x) \setminus \C(x)\,.
\end{align*}
With respect these sets $\C(x),\L(x)$, $x \in T$, we develop a series
of observations.
\begin{lemma}
  \label{lemma:atmostthree}
  If $y \in \C(x)$ then $x,y$ are twins. Consequently there can be at
  most two elements in $\C(x)$.
\end{lemma}
\begin{proof}
  If $y \in \C(x)$, then $y \lord x$ and $\height[T](y) \leq
  \height[T](x)$.  Since $y$ is above any lower cover of $x$, and
  distinct from such a lower cover, then $\height[T](x) \leq
  \height[T](y)$. It follows that $\height[T](x) = \height[T](y)$,
  hence if $z$ is a lower cover of $x$, then it is also a lower cover
  of $y$. Since $x,y$ have exactly one lower cover, it follows that
  $x,y$ are twins.  \qed
\end{proof}
\begin{lemma}
  \label{lemma:LandO}
  If $x,y$ are twins, then $\L(x) \subseteq O^{x,y}_{x}$. If $z \in
  \L(x)$ and $z ' \in O^{x,y}_{x}$ is such that $z' \leq z$, then $z'
  \in \L(x)$. That is, $\L(x)$ is a lower set of $O^{x,y}_{x}$.
\end{lemma}
\begin{proof}
  Let $z \in \L(x)$, so that $z \orth x$ and $z \orth \pi(x)$. The
  relation $y \leq z$ implies that $\pi(x) = \pi(y) \leq y \leq z$,
  and hence contradicts $z \orth \pi(x)$. Hence $y \not\leq z$ and $z
  \in O^{x,y}_{x}$. Let us suppose that $z' < z$ and $z' \orth x$.
  Then $\height(z') < \height(z)$ and $z' \lord z$, so that $z' \in
  \C(x)$. Since $z' \orth x$ then either $z' \orth \pi(x)$, or $\pi(x)
  \leq z'$. However, the latter property implies $\pi(x) \leq z$,
  which is not the case. Therefore $z' \orth \pi(x)$ and $z' \in
  \L(x)$.  \qed
\end{proof}

\begin{lemma}
  \label{lemma:Lempty}
  If $x,y,z$ are pairwise distinct twins, then $\L(x) = \emptyset$ and
  $z$ is the minimal element of $O^{x,y}_{x} \cap O^{x,y}_{y}$.  In
  particular $\O(x) = \C(x) \subseteq \set{y,z}$.
\end{lemma}
\begin{proof}
  By the previous observation $\L(x) \subseteq O^{x,y}_{x}$ and,
  similarly, $\L(x) \subseteq O^{x,z}_{x}$. Hence $\L(x) \subseteq
  O^{x,y}_{x} \cap O^{x,z}_{x} = O^{x,y,z}_{x} = \emptyset$, by Lemma
  \ref{lemma:twinsthree}.
  Since $x \orth z$ and $y \orth z$ then $z \in O^{x,y}_{x} \cap
  O^{x,y}_{y}$. If $z' < z$ then $z' < x$ and $z' < y$ hence $z'
  \not\in O^{x,y}_{x} \cup O^{x,y}_{y}$.   

  Finally, the relation $\C(x) \subseteq \set{y,z}$ follows from Lemma
  \ref{lemma:atmostthree}.
  \qed
\end{proof}

The previous observations motivate us to introduce the next
Definition.
\begin{definition}
  Let us say that $x,y \in T$ are a \emph{proper pair of twins} if
  they are distinct and $\set{z\,|\,\pi(z) = \pi(x)\,} = \set{x,y}$.
  We say that a linear order $\lord$ on $T$ is compatible with proper
  pair of twins if it satisfies \eqref{cond:height} and moreover
  \begin{align}
    \tag{TWINS}
    \label{cond:twins}
    O^{x,y}_{x} \supset O^{x,y}_{y}
    & \timplies x \lord y\,,
  \end{align}
  for each proper pair of twins $x,y$.
\end{definition}
Again is not difficult to see that such a linear order always exists
and in the following we shall assume that $\lord$ satisfies both 
 \eqref{cond:height} and \eqref{cond:twins}.

\vskip 12pt

We are ready to define a partial labelling of the event structure $\E$
whose domain is $T$.  W.r.t. $\lord$ let us say that $x \in T$ is
\emph{principal} if $\C(x) = \emptyset$.  Let $\Sigma =
\set{a_{0},a_{1},a_{2}}$ be a three elements totally ordered alphabet.
The labelling $\lambda : T \rTo \Sigma$ is defined by induction on
$\lord$ as follows:
\begin{enumerate}
\item If $x \in T$ is principal and $\height[T](x) = 0$, then we let
  $\lambda(x) = a_{0}$.
\item If $x \in T$ is principal and $\height[T](x) \geq 1$, let
  $\pi(x)$ be its unique lower cover. Since $\pi(x) \in T$ and $\pi(x)
  \lord x$, $\lambda(\pi(x))$ is defined and we let $\lambda(x) =
  \lambda(\pi(x))$.
\item If $x$ is not principal and $\L(x) = \emptyset$, then, by Lemma
  \ref{lemma:atmostthree}, we let $\lambda(x)$ be the least symbol not
  in $\lambda(\C(x))$.
\item If $x$ is not principal and $\L(x) \neq \emptyset$ then: 
  \begin{itemize}
  \item by Lemma \ref{lemma:Lempty} $\C(x) =
    \set{y}$ is a singleton and $x,y$ is a proper pair of twins,
  \item by Lemma \ref{lemma:LandO} $\L(x)$ is a lower set of
    $O^{x,y}_{x}$. By the condition \eqref{cond:twins}, $O^{x,y}_{x}
    \subseteq O^{x,y}_{y}$, so that $O^{x,y}_{x}$ is a linear order.
    Let therefore $z_{0}$ be the common least element of $\L(x)$ and
    $O^{x,y}_{x}$.
  \end{itemize} 
  We let $\lambda(x)$ be the unique symbol not in
  $\lambda(\set{y,z_{0}})$.
\end{enumerate}

\begin{proposition}
  \label{prop:labelling}
  For each $x,y \in T$, if $x \orth y$ then $\lambda(x) \neq
  \lambda(y)$.
\end{proposition}
\begin{proof}
  It suffices to prove that $\lambda(y) \neq \lambda(x)$ if $y \in
  \O(x)$. The statement is proved by induction on $\lord$. Let us
  suppose the statement is true for all $z \lord x$.

  (i) If $\height[T](x) = 0$ then $x$ is minimal in $T$, so that
  $\O(x) = \C(x)$. If moreover $x$ is principal then $\O(x) = \C(x) =
  \emptyset$, so that the statement holds trivially.

  (ii) If $x$ is principal and $\height[T](x) \geq 1$, then its unique
  lower cover $\pi(x)$ belongs to $T$.  Observe that $\O(x) = \L(x) =
  \set{y\in T\,|\,y \lord x \tand y \orth \pi(x)}$, so that if $y \in
  \O(x)$, then $y \orth \pi(x)$.  Since $y \lord x$ and $\pi(x) \lord
  x$, and either $y \in \O(\pi(x))$ or $\pi(x) \in \O(y)$, it follows
  that $\lambda(x) = \lambda(\pi(x)) \neq \lambda(y)$ from the
  inductive hypothesis.

  (iii) If $x$ is not principal and $\L(x) = \emptyset$, then $\O(x) =
  \C(x)$ and, by construction, $\lambda(y) \neq \lambda(x)$ whenever
  $y \in \O(x)$.

  (iv) If $x$ is not principal and $\L(x) \neq \emptyset$, then let
  $\C(x) = \set{y}$ and let $z_{0}$ be the common least element of
  $\L(x)$ and $O^{x,y}_{x}$.  Since by construction $\lambda(x) \neq
  \lambda(y)$, to prove that the statement holds for $x$, it is enough
  to pick $z \in \L(x)$ and argue that $\lambda(z) \neq \lambda(x)$.
  We claim that each element $z \in \L(x) \setminus \set{z_{0}}$ is
  principal.  If the claim holds, then $\lambda(z) = \lambda(\pi(z))$,
  so that $\lambda(z) = \lambda(z_{0})$ is inductively deduced.

  Suppose therefore that there exists $z \in \L(x)$ which is not
  principal and let $w \in \C(z)$. Observe that $x,y$ form a proper
  pair of twins, since otherwise $\L(x) = \emptyset$ by Lemma
  \ref{lemma:Lempty}. Similarly $w,z$ form a proper pair of twins:
  otherwise, if $z,w,u$ are pairwise distinct twins, then either $w
  \leq x$ or $u \leq x$ by Lemma \ref{lemma:twinsthree}. However this
  is not possible, since for example $z_{0} \leq \pi(x) < u \leq x$
  contradicts $z_{0} \orth x$.
  
  Since $y \lord x$, condition \eqref{cond:twins} implies $O^{x,y}_{x}
  \subseteq O^{x,y}_{y}$, and hence $z \in O^{x,y}_{x} \cap
  O^{x,y}_{y}$. If $w \in \C(z)$, then we cannot have $w \leq x$ or $w
  = y$, since again we would deduce $z_{0} \leq x$.  Thus Lemma
  \ref{lemma:Oinclusion} implies $O^{w,z}_{z} \supset O^{w,z}_{w}$.
  On the other hand, $w \lord z$ and condition \eqref{cond:twins}
  implies $O^{w,z}_{z} \subseteq O^{w,z}_{w}$.
  
  Thus, we have reached a contradiction by assuming $\C(z) \neq
  \emptyset$. It follows that $z$ is principal. \qed
\end{proof}

The obvious corollary of Proposition \ref{prop:labelling} is that if
$\E$ is already a sort of tree, then it has a nice labelling with $3$
letters. We state this fact as the following Theorem, after we having
made precise the meaning of the phrase ``$\E$ is a sort of tree.''
\begin{definition}
  Let us say that $\E$ is a \emph{forest} if every element has at most
  one lower cover. Let $\mathcal{F}_{3}$ be the class of event
  structures of degree $3$ that are forests.
\end{definition}
 
\begin{theorem}
   The labelling number of the class $\mathcal{F}_{3}$ is $3$.
\end{theorem}
As a matter of fact, let $\E$ be a forest, and consider the event
structure $\E_{\bot}$ obtained from $\E$ by adding a new bottom
element $\bot$. Remark that the graph $\G(\E_{\bot})$ is the same
graph as $\G(\E)$ apart from the fact that an isolated vertex $\bot$
has been added.  The set of events $E$ is a tree within $\E_{\bot}$,
hence the graph induced by $E$ in $\G(\E_{\bot})$ can be colored with
three colors. But this graph is exactly $\G(\E)$.


\section{More Upper Bounds}
\label{sec:w2}

The results presented in the previous section exemplify a remarkable
property of event structures of degree $3$: many types of subsets of
events induce a subgraph of $\G(\E)$ that can be colored with $3$
colors.  These include antichains by Corollary \ref{cor:antichain},
trees by Proposition \ref{prop:labelling}, and lower sets in $\CNF$,
that is configurations of $\E$.  As a matter fact, if $X \in \CNF$,
then $\width(X) \leq 3$, so that such a subset can be labeled with $3$
letters by Dilworth's Theorem.  Also, recall that the star of an event
$x \in E$ is the subgraph of $\G(\E)$ induced by the subset $\set{x}
\cup \set{y \in E\,|\,y \orth x}$. A star can also be labeled with $3$
letters.  To understand the reason, let $\mathcal{N}_{x}$ be the event
structure $\langle \set{y \,|\,y \orth x },\leq_{x},\conc_{x}
\rangle$, where $\leq_{x}$ and $\conc_{x}$ are the restrictions of the
causality and concurrency relations to the set of events of ${\cal
  N}_{x}$.
\begin{lemma}
  The degree of $\mathcal{N}_{x}$ is strictly less than $\deg(\E)$.
\end{lemma}
\begin{proof}
  The lemma follows since if $y \orth_{x} z$ in $\mathcal{N}_{x}$,
  then $y \orth z$ in $\E$.  As a matter of fact, let us suppose that
  $y \orth_{x} z$ in $\mathcal{N}_{x}$ and $y' < y$. If $y' \in
  \mathcal{N}_{x}$, then $y' \wconc z$. If $y' \not\in
  \mathcal{N}_{x}$, then $y' < x$.  It follows then from $z \orth x$
  and $y' < x$ that $y' \wconc x$. Similarly, if $z' < z$ then $z'
  \wconc x$.  \qed
\end{proof}
Hence, if $\deg(\E) = 3$, then $\deg(\mathcal{N}_{x}) \leq 2$, and it
can be labeled with $2$ letters, by \cite{rozoy}. It follows that star
of $x$ can be labeled with $3$ letters.

\breath

It might be asked whether this property can be exploited to construct
nice labellings. The positive answer comes from a standard technique
in graph theory \cite{zykov}. Consider a partition ${\cal P} =
\set{[z]\mid z \in E}$ of the set of events such that each equivalence
class $[z]$ has a labelling with $3$ letters. Define the quotient
graph $\G({\cal P},\E)$ as follows: its vertexes are the equivalence
classes of ${\cal P}$ and $[x] \orth \relax [y]$ if and only if there
exists $x' \in [x]$, $y' \in [y]$ such that $x' \orth y'$.
\begin{proposition}
  \label{prop:quotient}
  If the graph $\G({\cal P},\E)$ is $n$-coloriable, then $\E$ has a
  labelling with $3n$ colors. 
\end{proposition}
\begin{proof}
  For each equivalence class $[x]$ choose a labelling $\lambda_{[x]}$
  of $[x]$ with an alphabet with $3$ letters. Let $\lambda_{0}$ a
  coloring of the graph $\G({\cal P},\E)$ and define $\lambda(x) =
  (\lambda_{[x]}(x),\lambda_{0}([x]))$. Then $\lambda$ is a labelling
  of $\E$: if $x \orth y$ and $[x] = [y]$, then $\lambda_{[x]}(x) =
  \lambda_{[y]}(x) \neq \lambda_{[y]}(y)$ and otherwise, if $[x] \neq
  [y]$, then $[x] \orth\relax [y]$ so that $\lambda_{0}([x]) \neq
  \lambda_{0}([y])$.  \qed
\end{proof}
The reader should remark that Proposition \ref{prop:quotient}
generalizes Proposition \ref{prop:skew}. The Proposition also suggests
that a finite upper bound for the labelling number of event structures
of degree $3$ might indeed exist.

\breath

\noindent
\begin{minipage}[t]{0.50\linewidth}
  \mbox{\hspace{6pt}} 
  We conclude the paper by exemplifying how to use
  the Labelling Theorem on trees and the previous Lemma to construct a
  finite upper bound for the labelling number of event structures that
  we call simple due to their additional simplifying properties.
  Consider the event structure on the right and name it ${\cal S}$. In
  this picture we have used 
  dotted \hfill lines \hfill for\hfill the \hfill edges
  \hfill of\hfill the\hfill Hasse
\end{minipage}
\hfill
\begin{minipage}[t]{0.45\linewidth}
\vspace{-10mm}
$$
\xygraph{
  []*+{1}="1"([r]*+{2})
  [l(0.5)u]*+{3}="3"([r]*+{4}="4"[r]*+{5}="5")
  "3"[l(1)u]*+{6}="6"([r]*+{7}="7")
  "5"[l(0)u]*+{8}="8"([r]*+{9}="9")
  "1"(:@{.}"3",:@{.}"4")
  "2"(:@{.}"4",:@{.}"5")
  "3"(:@{.}"6",:@{.}"7")
  "5"(:@{.}"8",:@{.}"9")
  "6":@{=}"7"
  "8":@{=}"9"
  "3":@{=}"5"
  "6":@{-}"4"
  "7":@{-}"4"
  "8":@{-}"4"
  "9":@{-}"4"
  "7"[u(0.3)]="START"
  "8"[u(0.3)]="END"
  "START"[r(0.2)]="A"
  "4"[l(0.4)]="B"
  "4"[d(0.5)]="C"
  "4"[r(0.4)]="D"
  "END"[l(0.2)]="E"
  "START"
  -@`{"A","B","C","D","E"}
  "END"
}
$$  
\end{minipage}
diagram of $\langle E,\leq \rangle$, simple
lines for maximal concurrent pairs, and double lines for minimal
conflicts.  Concurrent pairs $x \conc y$ that are not maximal, i.e.
for which there exists $x',y'$ such that $x' \orth y'$ and either $x<
x'$ or $y < y'$, are not drawn. We leave the reader to verify that a
nice labelling of ${\cal S}$ needs at least $4$ letters. 
On the other hand, it shouldn't be difficult to see that a nice
labelling with $4$ letters exists. To obtain it, take apart events
with at most $1$ lower cover from the others, as suggested in the
picture.
\begin{longversion}
  below:
$$
\xygraph{ []*+{1}="1"([r]*+{2}) [l(0.5)u]*+{3}="3"(
  [rr]*+{5}="5") "3"[l(1)u]*+{6}="6"([r]*+{7}="7")
  "5"[l(0)u]*+{8}="8"([r]*+{9}="9")
  "1"(:@{.}"3",
  )
  "2"(
  :@{.}"5")
  "3"(:@{.}"6",:@{.}"7")
  "5"(:@{.}"8",:@{.}"9")
  "6":@{=}"7" "8":@{=}"9" "3":@{=}"5"
  "6":@{-}"2" "7":@{-}"2" "8":@{-}"1" "9":@{-}"1"
  "1"[r(3)u]*+{4}="4" }
$$
\end{longversion}
Use then the results of the previous section to label with three
letters the elements with at most one lower cover, and label the only
element with two lower covers with a forth letter. 

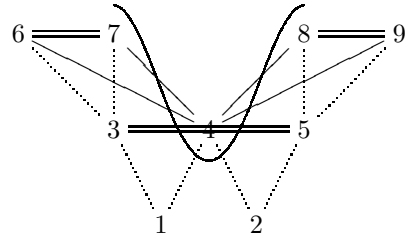
\begin{trash}
  To this goal, it is enough to split the event structures into a
  forest plus a residual, as suggested in figure \ref{fig:separation}.
  \begin{figure}
    \centering

$$
\xygraph{ []*+{1}="1"([r]*+{2}) [l(0.5)u]*+{3}="3"(
  [rr]*+{5}="5") "3"[l(1)u]*+{6}="6"([r]*+{7}="7")
  "5"[l(0)u]*+{8}="8"([r]*+{9}="9")
  "1"(:@{.}"3",
  )
  "2"(
  :@{.}"5")
  "3"(:@{.}"6",:@{.}"7")
  "5"(:@{.}"8",:@{.}"9")
  "6":@{=}"7" "8":@{=}"9" "3":@{=}"5"
  "6":@{-}"2" "7":@{-}"2" "8":@{-}"1" "9":@{-}"1"
  "1"[rr]*+{4}="4" } \hspace{5mm} \xygraph{ []*+{1}="1"([r]*+{2})
  [l(0.5)u]*+{3}="3"([r]*+{}="4"[r]*+{5}="5")
  "3"[l(0.5)u]*+{6}="6"([r]*+{}="7")
  "5"[l(0.5)u]*+{}="8"([r]*+{9}="9")
  "1"(:@{.}"3",
  )
  "2"(
  :@{.}"5")
  "3"(:@{.}"6",
  )
  "5"(
  :@{.}"9")
  "6":@{=}"5" "9":@{=}"3"
  "2"[r(1.5)]*+{4}="4" [l(0.5)u]*+{7}="7"[r]*+{8}="8"
  "4"(:@{.}"7",:@{.}"8") "7":@{=}"8" }
$$
\caption{Splitting $\E_{1}$ and $\E_{2}$}
\label{fig:separation}
\end{figure}
It will be enough to colour the forest and the residual with distinct
aphabets to obtain a nice labelling of the event structure.
\end{trash}

\begin{trash}
  To formalize these ideas, let $E_{1}$ be the set of all $x \in E$
  that have exactly one lower cover.  For $x \in E_{1}$ let $\rho(x) =
  \max \set{z \in E \,| z \leq x, z \not\in E_{1}}$.  The general
  technique presented in the next Proposition relies on standard graph
  coloring methods \cite{zykov}.
  \begin{proposition}
    \label{prop:G2}
    Let us suppose that $\rho(x) \not\comp \rho(y)$ whenever $x,y \in
    E_{1}$, $x \orth y$, and $\rho(x) \neq \rho(y)$.  Let
    $\G_{\geq2}(\E)$ be the subgraph of $\G(\E)$ induced by events
    with at least $2$ lower covers. Then $\gamma(\G(\E)) \leq 6
    \gamma(\G_{\geq2}(\E))$.
  \end{proposition}
  \begin{proof}
    As usual, we can suppose that $\E$ has just one minimal element
    $\bot$. Such a minimal element is an isolated element in $\G(\E)$,
    hence it does not matter to the goal of the coloring $\G(\E)$.

    Consider the graph $H$ having as vertices all the events, and
    where $\set{x,y}$ is an edge if and only $x \orth y$ and either
    $\set{x,y} \not\subseteq E_{1}$ or $\set{x,y} \subseteq E
    \setminus E_{1}$ but $\rho(x) \neq \rho(y)$. If $\tau : E
    \setminus E_{1} \rTo \Sigma$ is a coloring of the subgraph
    $\G(\E)$ induced by $E \setminus E_{1}$, then we define a function
    $\tau'$ of $H$ by letting $\tau'(x) = (\tau(\rho(x)),1)$ if $x \in
    E_{1}$, and $\tau'(y) = (\tau(y),0)$ otherwise. This is actually a
    coloring of $H$. To this goal, the only non trivial remark is that
    if $x,y \in E_{1}$, $x \orth y$ and $\rho(x)\neq \rho(y)$, then
    $\rho(x) \orth \rho(y)$.  Observe first that $\rho(x) \orth y$:
    otherwise, since $\rho(x) \leq x \orth y$, then $\rho(x) \leq y$
    and, by definition of $\rho$, $\rho(x) \leq \rho(y)$, a
    contradiction.  From $\rho(y) \orth y \geq \rho(y)$ we deduce
    either $\rho(x) \orth \rho(y)$ or $\rho(y) \leq \rho(x)$. Since
    the latter does not hold, then $\rho(x) \orth \rho(y)$.

    For each $y \in E\setminus E_{1}$ the set $T_{y} = \set{x \in
      E_{1} |\, \rho(x) = y}$ is a tree, and of course $T_{y} \cap
    T_{z} = \emptyset$ if $y \neq z$.
    Thus, for each $y \in E\setminus E_{1}$, let $\psi_{y} : T_{y}
    \rTo \set{b_{0},b_{1},b_{2}}$ be a nice labelling of $T_{y}$.
    Let $K$ be the graph having as vertices all the events, and where
    $\set{x,y}$ is an edge and either $\set{x,y} \subseteq E_{1}$,
    $\rho(x) = \rho(y)$ and $x \orth y$.  The function $\psi(x) =
    \psi_{\rho(x)}(x)$ is then a coloring of $K$.

    Since the each edge of $\G(\E)$ is either an edge of $H$ or an
    edge of $K$, then the function
    \begin{align*}
      \lambda(x) & = (\tau'(x),\psi(x))
    \end{align*}
    is a coloring of $\G(\E)$.  \qed
  \end{proof}
\end{trash}

\breath

A formalization of this intuitive method leads to the following
Definition and Proposition.
\begin{definition}
  We say that an event structure is \emph{simple} if 
  \begin{enumerate}
  \item it is graded, i.e.  $\height(x) = \height(y) - 1$ whenever $x
    \lcover y$,
  \item every size $3$ clique of $\G(\E)$ contains a minimal conflict.
  \end{enumerate}
\end{definition}
The event structure ${\cal S}$ is simple and proves that even simple
event structures cannot be labeled with just $3$ letters.
\begin{proposition}
  Every simple event structure of degree $3$ has a nice labelling with
  $12$ letters.
\end{proposition}
\begin{proof}
  Recall that $\fat(x)$ is the number of lower covers of $x$ and let
  $E_{n} = \set{ x \in E \mid \fat(x) = n}$. Observe that a simple
  $\E$ is such that $E_{3} = \emptyset$: if $x \in E_{3}$, then its
  three lower covers form a clique of concurrent events. Also, by
  considering the lifted event structure $\E_{\bot}$, introduced at
  the end of section \ref{sec:thetheorem}, we can assume that
  $\card(E_{0})=1$, i.e.  $\E$ has just one minimal element which
  necessarily is isolated in the graph $\G(\E)$.

  Let $\lord$ be a linear ordering of $E$ compatible with the height.
  W.r.t. this linear ordering we shall use a notation analogous to the
  one of the previous section: we let 
  \begin{align*}
    \O(x) & = \set{y \in E\,|\,y
      \lord x \tand y \orth x}\,,& 
    C(x) &  = \set{y \in E\mid y' \lcover y \timplies y' \lcover x}\,.
  \end{align*}
  \begin{claim}
    The subgraph of $\G(\E)$ induced by $E_{2}$ can be colored with
    $3$-colors.
  \end{claim}
  \takebreath We claim first that if $x \in E_{2}$ then $\O(x)
  \subseteq C(x)$. Let $y \in \O(x)$ and let $x_{1},x_{2}$ be the two
  lower covers of $x$.  From $x_{i} < x \orth y$ it follows $x_{i} <
  y$ or $x_{i} \conc y$.  If $x_{i} \conc y $ for $i =1,2$, then
  $y,x_{1},x_{2}$ is a clique of concurrent events. Therefore, at
  least one lower cover of $x$ is below $y$, let us say $x_{1} < y$.
  It follows that $\height(y) \geq \height(x)$, and since $y \lord x$
  implies $\height(y) \leq \height(x)$, then $x,y$ have the same
  height. We deduce that $x_{1} \lcover y$. If $y$ has a second lower
  cover $y'$ which is distinct from $x_{1}$, then $y',x_{1},x_{2}$ is
  a clique of concurrent events. Hence, if such $y'$ exists, then $y'
  = x_{2}$. Second, we remark that if $y,z \in C(x)$ and $x \in E_{2}$
  then $y \orth z$: if $y' < y$ then $y' \leq x$ so that $x \orth z$
  implies $y' \wconc z$, and symmetrically. It follows that for $x \in
  E_{2}$, $C(x)$ may have at most $2$ elements. In particular, the
  restriction of $\lord$ to $E_{2}$ is a $2$-elimination ordering.
  \eproofofclaim

  For $x \in E_{1}$ let $\rho(x) = \max \set{z \in E \mid z \leq x, z
    \not\in E_{1}}$ and $[x] = \set{y \in E_{1} \mid \rho(y) =
    \rho(x)}$.
\begin{shortversion}
     Let ${\cal P}$ be the partition
      $\set{E_{0}} \cup \set{[x] \mid x \in E_{1}} \cup
      \set{E_{2}}$.
\end{shortversion}
\begin{longversion}
    Define
    \begin{align*}
      {\cal P} & = \set{E_{0}} \cup \set{[x] \mid x \in E_{1}} \cup
      \set{E_{2}}\,.
    \end{align*}
\end{longversion}
  Since each $[x]$, $x \in E_{1}$, is a tree, the partition ${\cal P}$
  is such that each equivalence class induces a $3$-colorable subgraph
  of $\G(\E)$.
  
  \begin{claim}
    The graph $\G({\cal P},\E)$ is $4$-colorable.
  \end{claim}
  \takebreath
  Since $E_{0}$ is isolated in $\G({\cal P},\E)$, is it enough to
  prove that the subgraph of $\G({\cal P},\E)$ induced by the trees
  $\set{[x] \mid x \in E_{1} }$ is $3$-colorable.  Transport the
  linear ordering $\lord$ to a linear ordering on the set of trees:
  $[y] \lord [x]$ if and only if $\rho(y) \lord \rho(x)$.  Define
  $\O([x])$ as usual, we claim that $\O([x])$ may contain at most two
  trees.
  
  
  We define a function $f : \O([x]) \rTo C(\rho(x))$ as follows. If
  $[y] \orth\relax [x]$ and $[y] \lord [x]$ then we can pick $y' \in[y]$ and
  $x' \in [x]$ such that $y' \orth x'$. We notice also that $y' \orth
  \rho(x)$: from $\rho(x) \leq x' \orth y'$, we deduce $\rho(x) \orth
  y'$ or $\rho(x) \leq y'$. The latter, however, implies $\rho(x) \leq
  \rho(y)$, by the definition of $\rho$, and this relation contradicts
  $\rho(y) \lord \rho(x)$. Thus we let
  \begin{align*}
    f([y]) & = \min \set{z \mid \rho(y) \leq z \leq y' \tand z
      \not\leq \rho(x)}\,.
  \end{align*}
  By definition, $f([y]) \orth \rho(x)$ and every lower cover of
  $f([y])$ is a lower cover of $x$. This clearly holds if $f([y]) \neq
  \rho(y)$, and if $f([y]) = \rho(y)$ then it holds since $\rho(y)
  \lord \rho(x)$ implies $f([y]) \in \O(x) \subseteq C(x)$ by the
  previous Claim.  Thus the set $f(\O(x))$ has cardinality at most $2$
  and, moreover, we claim  that $f$ is injective.  Let us suppose that
  $f([y]) = f([z])$.  If $f([y]) = \rho(y)$, then $f([z]) = \rho(z)$
  as well and $[y] = [x]$.  Otherwise $f([y]) = f([x])$ impliees
  $\rho(y) = \rho(f([y])) = \rho(f([z])) = \rho(z)$ and $[y] = [z]$.
  \eproofofclaim
  
  Thus, by applying Proposition \ref{prop:quotient}, we deduce that
  $\G(\E)$ has a labelling with $12$ letters.  \qed
\end{proof}

 

\bibliographystyle{splncs}
\bibliography{biblio}

\begin{thebibliography}{10}

\bibitem{eventstructures1}
Nielsen, M., Plotkin, G.D., Winskel, G.:
\newblock Petri nets, event structures and domains, part {I}.
\newblock Theor. Comput. Sci. \textbf{13} (1981)  85--108

\bibitem{winskelnielsen}
Winskel, G., Nielsen, M.:
\newblock Models for concurrency.
\newblock In: Handbook of Logic and the Foundations of Computer Science.
  Volume~4.
\newblock Oxford University Press (1995)  1--148

\bibitem{winskelccs}
Winskel, G.:
\newblock Event structure semantics for {C}{C}{S} and related languages.
\newblock In Nielsen, M., Schmidt, E.M., eds.: ICALP. Volume 140 of Lecture
  Notes in Computer Science., Springer (1982)  561--576

\bibitem{varacca}
Varacca, D., Yoshida, N.:
\newblock Typed event structures and the {\it pi}-calculus: Extended abstract.
\newblock Electr. Notes Theor. Comput. Sci. \textbf{158} (2006)  373--397

\bibitem{faggian}
Faggian, C., Maurel, F.:
\newblock Ludics nets, a game model of concurrent interaction.
\newblock In: LICS, IEEE Computer Society (2005)  376--385

\bibitem{mellies}
Melli{\`e}s, P.A.:
\newblock Asynchronous games 2: The true concurrency of innocence.
\newblock In Gardner, P., Yoshida, N., eds.: CONCUR. Volume 3170 of Lecture
  Notes in Computer Science., Springer (2004)  448--465

\bibitem{bookoftraces}
Diekert, V., Rozenberg, G., eds.:
\newblock The book of traces.
\newblock World Scientific Publishing Co. Inc., River Edge, NJ (1995)

\bibitem{pratt}
Pratt, V.:
\newblock Modeling concurrency with partial orders.
\newblock Internat. J. Parallel Programming \textbf{15}(1) (1986)  33--71

\bibitem{rozoy}
Assous, M.R., Bouchitt{\'e}, V., Charretton, C., Rozoy, B.:
\newblock Finite labelling problem in event structures.
\newblock Theor. Comput. Sci. \textbf{123}(1) (1994)  9--19

\bibitem{arnold}
Arnold, A.:
\newblock An extension of the notions of traces and of asynchronous automata.
\newblock ITA \textbf{25} (1991)  355--396

\bibitem{thiagarajan}
Hoogers, P.W., Kleijn, H.C.M., Thiagarajan, P.S.:
\newblock An event structure semantics for general {P}etri nets.
\newblock Theoret. Comput. Sci. \textbf{153}(1-2) (1996)  129--170

\bibitem{dilworth}
Dilworth, R.P.:
\newblock A decomposition theorem for partially ordered sets.
\newblock Ann. of Math. (2) \textbf{51} (1950)  161--166

\bibitem{mycielski}
Mycielski, J.:
\newblock Sur le coloriage des graphs.
\newblock Colloq. Math. \textbf{3} (1955)  161--162

\bibitem{poem}
Niebert, P., Qu, H.:
\newblock The implementation of mazurkiewicz traces in poem.
\newblock In Graf, S., Zhang, W., eds.: ATVA. Volume 4218 of Lecture Notes in
  Computer Science., Springer (2006)  508--522

\bibitem{partialorderreduction}
Niebert, P., Huhn, M., Zennou, S., Lugiez, D.:
\newblock Local first search - a new paradigm for partial order reductions.
\newblock In Larsen, K.G., Nielsen, M., eds.: CONCUR. Volume 2154 of Lecture
  Notes in Computer Science., Springer (2001)  396--410

\bibitem{gratzer}
Gr{\"a}tzer, G.:
\newblock The congruences of a finite lattice.
\newblock Birkh\"auser Boston Inc., Boston, MA (2006) A proof-by-picture
  approach.

\bibitem{getco06}
Santocanale, L.:
\newblock Topological properties of event structures.
\newblock GETCO06 (August 2006)

\bibitem{fulkerson}
Fulkerson, D.R., Gross, O.A.:
\newblock Incidence matrices and interval graphs.
\newblock Pacific J. Math. \textbf{15} (1965)  835--855

\bibitem{zykov}
Zykov, A.A.:
\newblock On some properties of linear complexes.
\newblock Mat. Sbornik N.S. \textbf{24(66)} (1949)  163--188

\end{thebibliography}

\newpage
\appendix

\end{document}